\documentclass[11pt,letterpaper]{article}
\usepackage[T1]{fontenc}
\usepackage{lmodern}
\usepackage[margin=1in]{geometry}
\usepackage{amsmath,amssymb,amsthm,bm,graphicx,booktabs,microtype}
\usepackage[font=small,labelfont=bf]{caption}
\usepackage[numbers,sort&compress]{natbib}
\usepackage{chapterbib}
\usepackage[hidelinks]{hyperref}
\hypersetup{pdftitle={Residual Allan Stability under Periodic Auxiliary Ramsey Sampling},pdfauthor={Zhiqian Su}}
\usepackage{bookmark}
\newcommand{\Cov}{\operatorname{Cov}}

\makeatletter
\newcommand{\printreferences}{%
  \if@filesw\immediate\write\@auxout{\string\citation{apsrev42Control}}\fi
  \bibliographystyle{apsrev4-2}\bibliography{references,reference-style}}
\makeatother
\newtheorem{proposition}{Proposition}
\renewcommand{\thesection}{\Roman{section}}
\renewcommand{\thesubsection}{\thesection.\Alph{subsection}}

\begin{document}
\title{Residual Allan Stability under Periodic Auxiliary Ramsey Sampling}
\author{Zhiqian Su\\[3pt]
  \small Institute of Quantum Electronics, School of Electronics,\\
  \small Peking University, Beijing, China\\
  \small\href{mailto:zqsu@pku.org.cn}{zqsu@pku.org.cn}}
\date{}
\maketitle

\begin{abstract}
The timing of auxiliary Ramsey measurements can change the frequency stability associated with a drifting pulse shift at a fixed nominal Fisher-information rate for a constant shift. For periodic instantaneous observations of an Ornstein--Uhlenbeck shift with causal prediction between updates, we derive the exact residual covariance averaged over the update cycle and the corresponding Allan variance at arbitrary averaging times. Relative to a continuous Kalman reference at the same information rate, periodic estimation increases the cycle-averaged estimation-error variance, reduces the short-time residual Allan variance, and increases its long-time value. Its covariance decays more slowly at the origin and has a larger integral. We extend the result to instantaneous sensing of the sum of finitely many independent Ornstein--Uhlenbeck modes. Integrated observations preserve the short-time ordering for the scalar shift, and numerical calculations resolve the dependence of the crossover on window duration. A zero-order-hold comparison shows that the short-time ordering can change when the same estimator drives a different correction command. A three-level Ramsey model of coherent population trapping (CPT) maps the discriminator response to an effective information budget and illustrates how the dark times affect the choice of schedule.
\end{abstract}

\section{Introduction}

A Ramsey clock measures detuning through the phase accumulated between interrogating pulses \cite{Ramsey1950}. The pulses can also impose a phase shift through light shifts, amplitude errors, or interrogation transients. A constant pulse phase is equivalent to a frequency offset when a single dark time is used. Auto-balanced Ramsey (ABR) spectroscopy separates the two effects by comparing dark times, with a second servo parameter that corrects the pulse-dependent phase \cite{Sanner2018,Yudin2018PRApplied}. Combined-error-signal methods remove the common phase in a single discriminator \cite{Shuker2019}. Symmetric ABR sequences address preparation history in vapor-cell clocks \cite{HafizAPL2018,Hafiz2022,Rivera2025}. Pulsed CPT clocks implement related Ramsey interrogation \cite{Carle2021,HanCPTReview2024}, and cold-atom CPT provides another implementation of ABR \cite{LiColdCPT2024}. These techniques provide a way to sense a drifting pulse shift, in addition to the usual clock discriminator.

Auxiliary interrogation uses part of the available measurement time. One strategy allocates small blocks at frequent intervals. Another combines several successive measurements to estimate the shift more precisely at each update. Both can spend the same amount of wall time on auxiliary measurements. With white readout noise the two schedules then have the same Fisher information per unit time for a constant shift. Their performance for a time-dependent disturbance still depends on when the measurements arrive and on how the resulting correction varies between updates.

\begin{figure}[htbp]
 \centering
 \includegraphics[width=\textwidth]{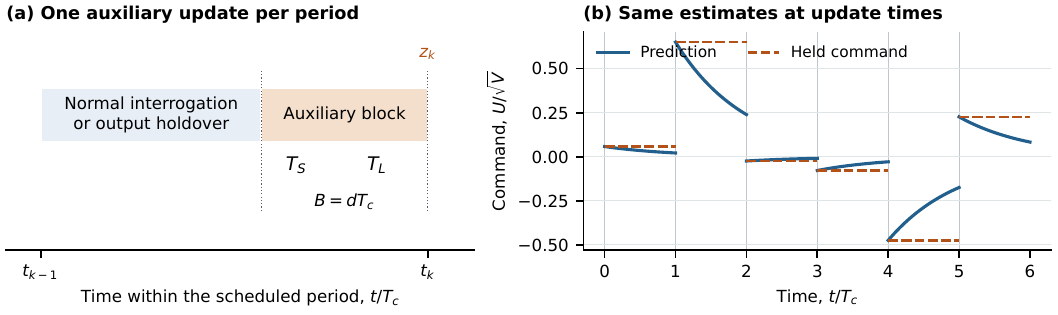}
 \caption{Auxiliary Ramsey sampling and idealized command paths. (a) One block combines the phase responses at dark times $T_S$ and $T_L$ to estimate the pulse-induced shift. The block occupies a fraction of each update cycle. (b) The periodic Kalman estimate jumps at readout times and decays according to the OU prediction between them. The zero-order-hold command uses the same update values through each cycle. The example trace is simulated at $\Xi=4$, $T_c=\tau_b$. Panel (a) is schematic.}
 \label{fig:scheme}
\end{figure}

Kalman filtering minimizes the mean-square estimation error for the matched linear Gaussian model \cite{Kalman1960,KalmanBucy1961}. Clock stability at averaging time $\tau$ instead depends on a difference of adjacent averages \cite{Allan1966,Riley2008}. This difference responds to fluctuations over a range of timescales, so the covariance of the error across an interval matters alongside its variance at a single instant. Allan analysis helps identify clock disturbances and assess Kalman estimates \cite{Wu2019,Wu2023}. Missing records in clock comparisons also require a suitable Allan estimator \cite{GalleaniSesia2019}. Clock-ensemble estimation can favor different choices at long and short averaging times \cite{Ishizaki2026}. The question here is how distributing one auxiliary measurement budget in time changes the covariance and Allan variance of an identifiable pulse shift.

Optical time-scale simulations quantify the cost of intermittent reference-clock operation \cite{Milner2019}, and steering experiments compare how often a frequency correction is applied \cite{Godel2019}. Rubidium-clock steering also uses adaptive noise identification \cite{LiuRb2024}. Resource-constrained sensor scheduling compares continuous allocations with periodic schedules, including for mean-square tracking \cite{LeNy2011}. With random measurement arrivals, filtering stability also depends on the observation process \cite{Marelli2019}. Enzer \emph{et al.} use the Allan deviation of per-update synthesizer corrections to diagnose disturbances in an ion clock, distinguishing that series from the clock output \cite{Enzer2021}. We analyze the unremoved shift $b-\widehat b$ between measurements. A further timing effect is the Dick effect, where the periodic interrogation sensitivity aliases local-oscillator frequency noise into the sampled clock readout \cite{Santarelli1998,Danet2014}. Here the pulse shift is the stochastic input, and auxiliary updates shape its residual covariance.

We obtain the phase-averaged covariance and Allan variance of a continuously predicted OU residual under periodic measurements. This yields an analytic reversal between periodic and continuous observation at equal nominal information rate. The covariance at zero measures the mean-square error, its slope controls the short-time Allan variance, and its area controls the long-time Allan variance. We then extend the ordering to multiple independent drift scales, examine duration and command-hold effects, and connect the measurement strength to a two-dark-time CPT sequence.

\section{Physical and stochastic model}
\label{sec:model}

A pulse-dependent phase shifts the apparent center of a Ramsey fringe. We denote by $b(t)$ its equivalent fractional-frequency shift in the short-dark-time channel, calibrated in Sec.~\ref{sec:ramsey}. Take $b$ to be a stationary Ornstein--Uhlenbeck (OU) process,
\begin{equation}
 db=-b\,dt/\tau_b+\sqrt{2V/\tau_b}\,dW,
\end{equation}
where $V$ is the stationary variance and $\tau_b$ the correlation time. This one-pole spectrum describes a mean-reverting disturbance \cite{Bibbona2008}. Time-scale models also use power-law noise and drift components \cite{Wu2019,Wu2023,Riley2008}. Section~\ref{sec:multimode} treats independent OU modes with several correlation times.

An auxiliary block of duration $B=dT_c$ ends at $t_k=kT_c$. The effective point observation, $z_k=b(t_k)+\epsilon_k$, has independent Gaussian errors of variance $R_{\rm blk}=S_m/B$ and $0<d\leq1$. The point model places the full block's measurement information at its endpoint and uses the endpoint shift as the signal. Section~\ref{sec:window} retains the signal variation during acquisition. Here $S_m$ is the full-duty white measurement noise intensity, in squared fractional frequency times seconds. The observation budget is the Fisher information per unit time for a constant shift,
\begin{equation}
 {\cal J}=(T_cR_{\rm blk})^{-1}=d/S_m .
 \label{eq:nominal}
\end{equation}
This convention compares equally precise estimates of a constant shift over equal total time. Information about a stochastic trajectory still depends on when observations arrive. In the continuous benchmark, $dY=b(t)dt+\sqrt{S_m/d}\,dW_m$ with $W_m$ independent of $W$. Over time $T$, the increment of $Y$ has Fisher information $T{\cal J}$ for a constant shift. This is an equal-resource reference for the sampled Ramsey sequence \cite{KalmanBucy1961,LeNy2011}.

Table~\ref{tab:scales} gives the time and variance scales used below.

\begin{table}[htbp]
\caption{\label{tab:scales}Normalized variables for one OU mode. For mixtures use $\tau_{\rm ref}$ in place of $\tau_b$.}
\centering
\begin{tabular}{ll}
\toprule
Symbol & Meaning\\
\midrule
$\eta=T_c/\tau_b$ & Update interval\\
$d=B/T_c$ & Auxiliary wall-time fraction\\
$\Xi={\cal J}V\tau_b$ & Information per OU timescale\\
$r=R_{\rm blk}/V=(\Xi\eta)^{-1}$ & Block noise variance\\
$u=(t-t_k)/\tau_b$ & Time since last update\\
$x=\tau/\tau_b$ & Allan averaging time \\
\bottomrule
\end{tabular}

\end{table}

The estimate jumps at each update, then follows the conditional mean $\widehat b(t_k+s)=e^{-s/\tau_b}\widehat b(t_k^+)$. The primary observable is the estimation residual $e=b-\widehat b$. If an actuator applies $U(t)=\widehat b(t)$, its pulse-shift residual $b-U$ equals $e$. A command held at $\widehat b(t_k^+)$ through a cycle gives a different residual. Figure~\ref{fig:scheme} shows both command paths with the same observations. Section~\ref{sec:actuation} compares their Allan statistics.

\section{Exact residual covariance and Allan variance}
\label{sec:dynamic}

\subsection{Periodic state estimation}

Normalize times by $\tau_b$ and variances by $V$. The OU transition coefficient is $a=e^{-\eta}$ and its process variance is $1-a^2$. In steady periodic operation the prior and posterior variances are \cite{Kalman1960,Bittanti1990,DeNicolaoStrada1998}
\begin{align}
 p^- &=a^2p^++1-a^2,& p^+&=(1-K)p^-,\\
 K&=\frac{p^-}{p^-+r}.&&
\end{align}
Equivalently $p^+$ solves $a^2(p^+)^2+(1-a^2)(1+r)p^+-r(1-a^2)=0$. Its negative constant term and positive leading coefficient select a unique positive root, which lies below unity. The variance then recovers through the cycle,
\begin{equation}
 p(u)=1+(p^+-1)e^{-2u},\qquad
 \bar p=1+(p^+-1)\frac{1-a^2}{2\eta}.
\end{equation}
The cycle-averaged mean-square error is $V\bar p$.

\begin{figure}[htbp]
 \centering
 \includegraphics[width=\textwidth]{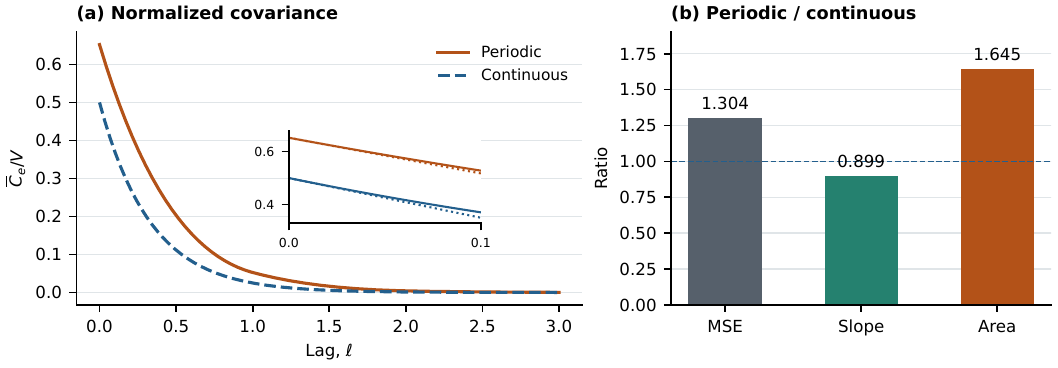}
 \caption{Origin of the stability reversal for a scalar OU pulse shift at $\Xi=4$ and $T_c=\tau_b$. (a) Normalized residual covariances and the initial tangents (inset). (b) Ratios of the zero-lag value (cycle MSE), the magnitude of the initial slope, and the total covariance area, all relative to the matched continuous reference. The slope and area ratios are the short- and long-time limits of the Allan-variance ratio. Curves and coefficients are computed from the exact covariances.}
 \label{fig:covariance}
\end{figure}

\subsection{Covariance across updates}

At any origin phase $u$, the current estimation error propagates to a lag $\ell$ through OU decay and through multiplication by $\alpha=1-K$ at each update. Future process and readout noises are independent of the error at that origin. If $\ell=m\eta+s$ with $0\le s<\eta$, origins before $\eta-s$ encounter $m$ updates and the remaining origins encounter $m+1$. Averaging uniformly over the origin within the cycle gives
\begin{equation}
\begin{split}
 c_\eta(\ell)&=\overline C_e(\ell\tau_b)/V\\
 &=\frac{e^{-\ell}\alpha^m}{\eta}
 \left[\int_0^{\eta-s}p(u)du+
 \alpha\int_{\eta-s}^{\eta}p(u)du\right].
\end{split}
\label{eq:sampledcov}
\end{equation}
In particular $c_\eta(0)=\bar p$. The covariance factors as $(a\alpha)^m f(s)$, with elementary $f$. Section~\ref{S:sec:point} gives its moments by geometric summation. Uniform averaging over one cycle gives a single lag covariance. A time-local statistic retains the dependence on the origin \cite{GalleaniTavella2008}.

\subsection{Time averages of the residual}

\textit{Phase-averaged Allan variance.} Let $X_1,X_2$ be adjacent averages of the continuous residual over intervals of duration $\tau$. The two-sample Allan variance is $\sigma_{A,e}^2=\tfrac12\mathbb E[(X_2-X_1)^2]$ \cite{Allan1966,Riley2008}. Here the expectation includes the noise and a starting phase drawn uniformly over one update cycle. Sampling the clock at a fixed phase supplies a different sequence for Allan analysis \cite{Makdissi2010}. The phase-averaged variances of $X_1$ and $X_2$ are equal. The Allan variance is twice the variance of $X_1$ minus twice that of $(X_1+X_2)/2$. Integrating the covariance for those two window lengths gives
\begin{align}
 \frac{\sigma_{A,e}^2}{V}
 &=\frac{1}{x^2}\int_0^{2x}k_x(\ell)c_\eta(\ell)\,d\ell,\label{eq:allan-kernel}\\
 k_x(\ell)&=4(x-\ell)_+-(2x-\ell)_+,\nonumber
\end{align}
with $(z)_+=\max(z,0)$. It is applicable to any $\tau/T_c$ on the continuous residual.

The kernel has zero area and first moment $-2x^3/3$. The constant covariance term therefore cancels at short $\tau$, leaving the right derivative at zero. At long $\tau$, lags within the covariance decay time lie on the branch $k_x(\ell)=2x-3\ell$. Exponential decay makes the covariance area and its first moment finite, giving
\begin{equation}
 \frac{\sigma_{A,e}^2}{V}=
 \begin{cases}
 -\tfrac23 c_\eta'(0^+)x+O(x^2), & x\to0,\\[2pt]
 \tfrac2x\int_0^\infty c_\eta(\ell)d\ell+O(x^{-2}), & x\to\infty .
 \end{cases}
 \label{eq:allan-limits}
\end{equation}
Thus the value at zero, the initial slope, and the area set three distinct measures of the same estimation error.

\section{Matched-rate stability reversal}

For the continuous reference, $\Xi p_c^2+2p_c-2=0$, so
\begin{equation}
 p_c=\frac{2}{1+\sqrt{1+2\Xi}},\qquad
 c_c(\ell)=p_c e^{-(2/p_c-1)\ell}.
\end{equation}
Equation~\eqref{eq:allan-kernel} applied to $c_c$ gives the stationary exponential Allan curve \cite{Bibbona2008,Goerlich2021}.

\begin{proposition}[Short-to-long reversal]
For stationary scalar OU drift at finite $\eta>0$ and $\Xi>0$, assume independent readout noise and periodic and continuous steady state. Periodic instantaneous observation with causal OU prediction gives $\bar p>p_c$. Its phase-averaged residual Allan variance is lower than that of the continuous reference for sufficiently short averaging times and higher for sufficiently long times. The two curves cross at least once.
\end{proposition}

\subsection{Information and mean-square error}

Write $w=1/p$. Between updates $w'=2w-2w^2$, and the update adds $\Xi\eta$ to the precision. Over one cycle this gives $\langle w^2\rangle-\langle w\rangle=\Xi/2$. The fluctuations of $w$ make $\langle w\rangle^2-\langle w\rangle<\Xi/2$. Because $w\ge1$, comparison with the continuous precision yields
\begin{equation}
 \langle w\rangle<p_c^{-1},\qquad
 \bar p=\langle w^{-1}\rangle\ge\langle w\rangle^{-1}>p_c .
 \label{eq:jensen}
\end{equation}
Precision acquired in jumps is lost most rapidly just after each jump, when precision is highest.

\subsection{Slope, area, and crossing}

Differentiating Eq.~\eqref{eq:sampledcov} at zero gives $-c_\eta'(0^+)=\bar p+Kp^-/\eta$. The second term comes from the fraction of short windows straddling an update. Its numerator $Kp^-=p^--p^+$ is the variance removed there and the normalized variance of the correction increment. The variance regained between observations is $2\eta(1-\bar p)$. This gives
\begin{equation}
 -c_\eta'(0^+)=2-\bar p<2-p_c=-c_c'(0^+).
 \label{eq:slopes}
\end{equation}
The periodic estimate removes less variance per unit time and adds less short-time variation through its update increments.

For the area, integrating $(\ln p)'=2(w-1)$ over a cycle gives $\ln\alpha=-2\eta(\langle w\rangle-1)$. The number $N$ of updates encountered by a lag $\ell$ has mean $\ell/\eta$. Jensen's inequality applied to $c_\eta=e^{-\ell}\langle\exp(\ln p+N\ln\alpha)\rangle$, followed by $\langle\ln p\rangle\ge-\ln\langle w\rangle$, yields
\begin{equation}
 c_\eta(\ell)\ge\frac{\exp[-(2\langle w\rangle-1)\ell]}{\langle w\rangle}
 >c_c(\ell),\qquad \ell\ge0.
 \label{eq:areaorder}
\end{equation}
Hence the periodic covariance retains a larger area even though it initially decreases more slowly. Equation~\eqref{eq:allan-limits} gives opposite Allan orderings in the two limits. Continuity for $\tau>0$ gives the crossing.

\subsection{Numerical crossing map}

\begin{figure}[htbp]
 \centering
 \includegraphics[width=\textwidth]{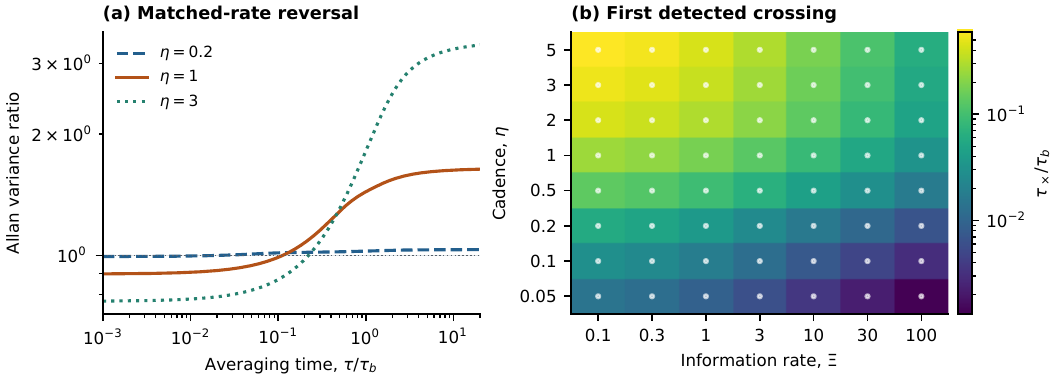}
 \caption{Crossover between residual Allan curves at matched information rate. (a) Allan-variance ratios at $\Xi=4$ for three normalized update intervals. Unity marks equality with the continuous reference. (b) First detected crossing, normalized by $\tau_b$, for 56 point-observation cases. Dots mark evaluated pairs. The curves are numerical evaluations of the exact Allan kernel. The grid detects sign changes over $10^{-5}\leq \tau/\tau_b\leq100$.}
 \label{fig:crossing}
\end{figure}

At $\Xi=4$ and $T_c=\tau_b$, the periodic and continuous covariances have normalized variances $0.6521$ and $0.5000$, slope magnitudes $1.348$ and $1.500$, and areas $0.2741$ and $1/6$ [Fig.~\ref{fig:covariance}]. Their Allan deviations cross near $0.113\tau_b$, with ratios $0.994$, $1.199$, and $1.276$ at $\tau/\tau_b=0.1,1,10$ [Fig.~\ref{fig:crossing}(a)].

On the grid of 56 $(\Xi,\eta)$ pairs in Fig.~\ref{fig:crossing}(b), a search over $10^{-5}\le\tau/\tau_b\le 100$ detected a single sign change at each pair. The first detected crossing varies from $0.0013\tau_b$ to $0.591\tau_b$ on that grid. The spread in crossing times quantifies how strongly the relevant averaging time depends on cadence and information rate.

\section{Extensions and implementation dependence}
\subsection{A finite integration window}
\label{sec:window}

We replace the point observation by the trailing average
\begin{equation}
 z_k=B^{-1}\int_{t_k-B}^{t_k} b(t)\,dt+\epsilon_k .
 \label{eq:window-observation}
\end{equation} A Ramsey sequence with a resolved sensitivity profile can instead weight the integrand by a normalized function $w_k(t)$. We use a uniform boxcar sensitivity over the integration window. Its weighting differs from the sequence-dependent CPT response shown in Fig.~\ref{fig:ramsey}(b). The update occurs at its end, after the full readout has arrived. The integrated state and its endpoint share process noise \cite{LjungWills2010}. Time aggregation also enters trajectory inference from interval observations \cite{FoliaRattray2018}.

In normalized units, write $\beta=B/\tau_b$. Conditional on the data from preceding cycles, the covariance on a cycle is $C_0(u,v)=e^{-|u-v|}+(p^+-1)e^{-(u+v)}$. It gives an innovation variance $D$ equal to its double window integral divided by $\beta^2$ plus $r$. Its covariance $M$ with the endpoint error is the single window integral of $C_0(\eta,v)$ divided by $\beta$. The gain is $K=M/D$, and the posterior variance is $p^+=p^--M^2/D$ with $p^-=C_0(\eta,\eta)$. To propagate across an update from a phase earlier in the same cycle, the covariance with the innovation is likewise integrated over the window. Section~\ref{S:sec:window} gives the full lag covariance.

The nominal budget in Eq.~\eqref{eq:nominal} is still the Fisher information for a constant shift, since the normalized window leaves a constant unchanged. In the window sweep we fix $\eta$ and $\Xi$ and vary $B/T_c$, so $R_{\rm blk}$ is held fixed. The comparison adjusts $S_m=R_{\rm blk}B$ with $B$. For fixed $S_m$ and duty $d$, changing $T_c$ changes both $B=dT_c$ and the block variance, as in Sec.~\ref{sec:ramsey}.

At a common prior, an average over a nonzero window is less informative about the endpoint than a direct observation of that endpoint with the same independent readout noise. The covariance $M$ is at most $p^-$, and the remaining conditional observation variance is at least $r$. Monotonicity of the scalar Riccati map therefore gives a larger steady cycle MSE than in the point model. The initial covariance slope still obeys Eq.~\eqref{eq:slopes} with the window-specific $\bar p$: at the update, $KM=M^2/D$ equals the variance drop. Short-time Allan variance thus remains lower than in the continuous reference at matched rate.

\begin{figure}[htbp]
 \centering
 \includegraphics[width=\textwidth]{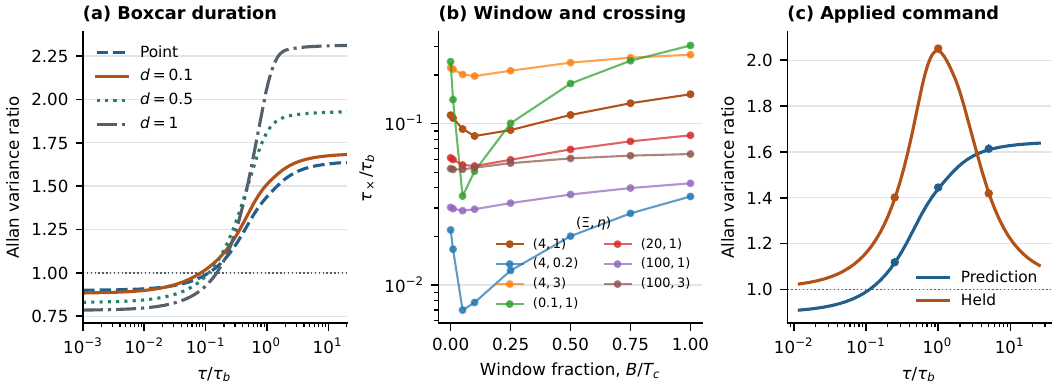}
 \caption{Observation duration and actuation. (a) Ratios of periodic to matched continuous residual Allan variance in the scalar OU model at $\Xi=4$ and $\eta=1$, for several boxcar widths $B/T_c$. (b) Numerically detected crossover times at the seven cadences and information settings of the window sweep, including the instantaneous limit. (c) Allan-variance ratios to the matched continuous reference for prediction and hold, using the same instantaneous samples ($\Xi=4$, $\eta=1$). Points give Monte Carlo estimates divided by the analytic continuous reference. Bars show pointwise 95\% intervals across independent trajectories.}
 \label{fig:windows}
\end{figure}

At $\Xi=4$ and $\eta=1$, the first crossing occurs at $\tau/\tau_b=0.113$ for point observation and in the range $0.084$--$0.152$ for the seven finite windows in Fig.~\ref{fig:windows}. At $\tau=0.1\tau_b$ the Allan-variance ratio changes from $0.989$ at zero width to $1.018$ at $B=0.1T_c$. Here a window of one tenth of a cycle switches which schedule gives the smaller variance at the target averaging time. For the 49 positive-window parameter combinations, the calculated covariance area exceeds that of the matched continuous reference.

\subsection{Independent OU modes with different correlation times}
\label{sec:multimode}

A drifting pulse shift can have distinct spectral corners. Let $b=\bm h^T\bm b$ be the sum of independent stationary OU components, with $\bm h$ a vector of ones, variances $V_i>0$, and correlation times $\tau_i>0$. We compare spectra at fixed total variance $V=\sum_i V_i$ and fixed $\tau_{\rm ref}=\sum_i V_i\tau_i/V$. These fix both the zero-lag covariance and the uncorrected covariance area, $C_b(0)=V$ and $\int_0^\infty C_b(t)\,dt=V\tau_{\rm ref}$. There is no single decay time. In units of $\tau_{\rm ref}$ and $V$, write
\begin{equation}
 D=\operatorname{diag}(\tau_{\rm ref}/\tau_i),\qquad
 \Sigma=\operatorname{diag}(V_i/V),\qquad Q=2D\Sigma.
 \label{eq:multimode-state}
\end{equation}
The vector drift is $-D$ with diffusion covariance $Q$. An instantaneous observation senses the sum with noise variance $r=(\Xi\eta)^{-1}$, where $\Xi=dV\tau_{\rm ref}/S_m$ and $\eta=T_c/\tau_{\rm ref}$. The continuous filter observes the same sum with white-noise intensity $1/\Xi$.

The mode estimation errors become correlated because the modes share the same observation. If $P(u)$ is the vector error covariance at cycle phase $u$ and $G(u+\ell,u)$ propagates the error across the intervening drifts and updates, the exact periodic residual covariance of the sum is
\begin{equation}
 c_p(\ell)=\frac{1}{\eta}\int_0^\eta
 \bm h^TG(u+\ell,u)P(u)\bm h\,du .
 \label{eq:multimode-covariance}
\end{equation}
The same adjacent-average kernel as for a scalar shift converts $c_p$ to Allan variance. Section~\ref{S:sec:multimode} gives the matrix propagator and its continuous counterpart.

\begin{proposition}[Matched-rate ordering for OU mixtures]
\label{prop:multimode}
Consider a finite independent OU mixture, instantaneous observations of its sum with independent Gaussian readout noise, and steady-state Kalman filters. Periodic observation at finite $\eta>0$ with causal prediction has strictly greater cycle-averaged MSE of the summed shift than continuous observation at the same $\Xi>0$. The phase-averaged periodic residual Allan variance is lower for sufficiently short averaging times and higher for sufficiently long averaging times. The curves therefore cross at least once.
\end{proposition}

The two limits are determined by different features of the residual covariance. Between updates, the precision matrix $W=P^{-1}$ obeys $W'=DW+WD-WQW$. Its jump is $\Xi\eta\,\bm h\bm h^T$. Averaging the precision equation retains the positive semidefinite contribution $\langle(W-\bar W)Q(W-\bar W)\rangle$ to the continuous precision equation evaluated at $\bar W$. Stability of the continuous filter implies $\bar W\preceq P_c^{-1}$, where $P_c$ is the continuous error covariance. Convexity of matrix inversion then gives $\bar P\succeq P_c$, with strict inequality in the measured direction since $P(u)\bm h$ varies through the cycle.

The initial covariance slope receives contributions from diffusion and the measurement innovations. For $\bm q=P^-\bm h$, define $\tilde{\bm q}=\bm q/\sqrt{1+\Xi\eta\,\bm h^T\bm q}$ and $\bm q_c=P_c\bm h$. The slope magnitudes in periodic and continuous tracking are, respectively,
\begin{align}
 -c_p'(0^+)&=\tfrac12\bm h^TQ\bm h
              +\tfrac{\Xi}{2}(\bm h^T\tilde{\bm q})^2,\nonumber\\
 -c_c'(0^+)&=\tfrac12\bm h^TQ\bm h
              +\tfrac{\Xi}{2}(\bm h^T\bm q_c)^2.
 \label{eq:multimode-cusp}
\end{align}
The covariance balance constrains each $|\tilde q_i|$ by the corresponding positive $q_{c,i}$, strictly for at least one mode. The innovations therefore contribute less to the periodic cusp than to the continuous one.

In normalized time, the two-sided spectrum at zero equals twice the covariance area. With $a_i=D_{ii}$, the dimensionless uncorrected spectrum at zero is $S_0=2\sum_i\Sigma_{ii}/a_i$. The point-sampled spectrum, scaled by $\eta$, is $S_a=\eta\sum_i\Sigma_{ii}\coth(a_i\eta/2)>S_0$. If $g_p$ is the gain from the sampled readout to the cycle-averaged estimate at zero frequency, the long-time Allan coefficient is
\begin{equation}
 L_p=S_0(1-2g_p)+(S_a+1/\Xi)g_p^2.
 \label{eq:multimode-long}
\end{equation}
The corresponding continuous expression uses $S_0$ and gain $g_c$. The componentwise covariance bound and the increased readout spectrum give $0<g_p<g_c<S_0/(S_0+1/\Xi)$. The quadratic with $S_a$ replaced by $S_0$ decreases on this gain interval. Consequently $L_p>L_c$. Periodic sampling thus changes the long-time residual through both the sampled shift spectrum and the gain applied to it. Section~\ref{S:sec:mixture-ordering} derives the gain bounds.

\begin{figure}[htbp]
 \centering
 \includegraphics[width=\textwidth]{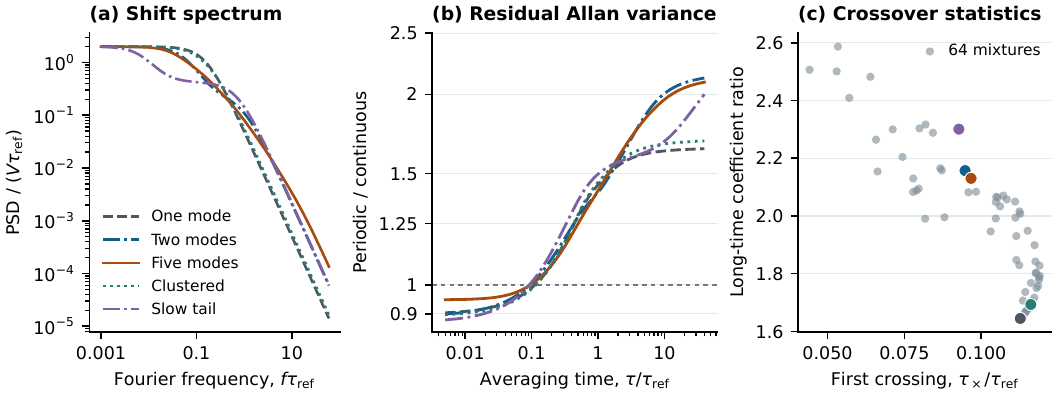}
 \caption{Instantaneous sampling of independent OU mixtures at $\Xi=4$ and $T_c=\tau_{\rm ref}$, with common variance $V$ and integrated correlation time $\tau_{\rm ref}$. (a) Two-sided PSD of the uncorrected shift, for Fourier frequency $f$ in cycles per unit time. (b) Ratio of periodic to continuous residual Allan variances, each evaluated for the same shift spectrum. (c) First sign-changing crossing located on the numerical scan and the ratio of the long-time Allan-variance coefficients. Colored points use the spectra in panel (a), and gray points denote the 64 seeded mixtures. The curves and ratios use the phase-averaged matrix covariance.}
 \label{fig:multimode}
\end{figure}

Figure~\ref{fig:multimode} compares several spectra with the same total variance $V$ and integrated correlation time $\tau_{\rm ref}$. At $\Xi=4$ and $\eta=1$, the two- and five-mode examples cross at $0.0947\tau_{\rm ref}$ and $0.0967\tau_{\rm ref}$, respectively, compared with $0.113\tau_{\rm ref}$ for a single mode. A small slow component can have a pronounced effect on long-time stability. Assigning it $4\%$ of the variance raises the long-time periodic-to-continuous Allan-variance coefficient ratio to $2.30$. Across 64 additional mixtures, the first detected crossing ranges from $0.0442\tau_{\rm ref}$ to $0.119\tau_{\rm ref}$.

\subsection{Prediction versus held correction}
\label{sec:actuation}

Figure~\ref{fig:windows}(c) shows how a held command changes the short-time ordering in the example. With continuous prediction, the applied command follows the conditional mean throughout the cycle. A zero-order hold uses the same observations and keeps the posterior estimate $U(t)=\widehat b(t_k^+)$ until the next update. Its residual is
\begin{equation}
 b(t)-U(t)=e(t)+(e^{-u}-1)\widehat b(t_k^+),\qquad
 u=(t-t_k)/\tau_b .
\end{equation}
Their cross-covariance must also be retained at nonzero lag. We compute it by propagating the joint state $(b(t_k),\widehat b(t_k^+))$ through the cycle and through subsequent updates (Sec.~\ref{S:sec:hold}).

The short-time effect can be seen from the command jumps. With prediction-based actuation, the normalized jump variance is $Kp^-$. With a hold, it is $Kp^-+(1-a)^2(1-p^+)$. The extra term comes from the difference between the old held command and the decayed estimate just before the measurement. It adds in variance because the new innovation is orthogonal to past measurements. The short-time Allan coefficient equals $2/3$ from the normalized OU diffusion plus the command jump variance divided by $3\eta$. At $\Xi=4$ and $\eta=1$, the resulting coefficient relative to continuous observation is $0.899$ for prediction and $1.006$ for a hold. The corresponding long-time ratios computed from the joint covariance are $1.645$ and $1.032$. In this case, the held residual has a larger short-time and long-time coefficient than the continuous reference. The held command is refreshed at each update, whereas the predicted command evolves continuously between updates.

\section{From CPT Ramsey readout to a correction schedule}
\label{sec:ramsey}

\subsection{Ideal phase-domain estimate}

Near the central fringe, the ideal phase model gives demodulated, contrast-normalized readouts $x_j=\omega T_j+\phi+n_j$ for $j=S,L$. The common pulse phase $\phi$ and residual angular detuning $\omega$ enter differently as $T_j$ changes, as used in auto-balanced Ramsey interrogation \cite{Sanner2018,Rivera2025}. With $R_\phi=\Cov[(n_S,n_L)^T]$ and $\bm v=(T_L,-T_S)^T$, the pulse-phase estimate and its single-sequence variance in equivalent fractional-frequency units are
\begin{align}
 \widehat{\phi}&=\frac{T_Lx_S-T_Sx_L}{T_L-T_S}, \label{eq:phase-estimate}\\
 R_1&=\frac{\bm v^T R_\phi\bm v}{[\omega_0 T_S(T_L-T_S)]^2}. \label{eq:aux-noise}
\end{align}
Dividing $\widehat\phi$ by $\omega_0T_S$, with $\omega_0$ the carrier angular frequency, gives the equivalent fractional-frequency shift of the short-time discriminator. The matrix $R_\phi$ allows correlated readouts. Quantum projection noise sets a readout scale for state-population measurements \cite{Itano1993}. The atomic error signal can also be used to infer local-oscillator noise and optimize the servo parameters \cite{Leroux2017}.

\begin{table}[htbp]
 \centering
 \caption{Parameters of the CPT example. The pulse duration, detection delay and reference dark times follow Ref.~\cite{Rivera2025}. The optical rates, discriminator noise and drift are model inputs. The cesium carrier is $\nu_0=9.192631770~\mathrm{GHz}$.}
 \label{tab:cpt-parameters}
 \begin{tabular}{@{}ll@{}}
 \toprule
 Parameter & Value \\
 \midrule
 $\Gamma_e/(2\pi),\ \Omega/(2\pi)$ & $4.6~\mathrm{MHz},\ 0.30~\mathrm{MHz}$ \\
 $\gamma_g/(2\pi),\ \Delta/(2\pi)$ & $0.80~\mathrm{kHz},\ 0$ \\
 Pulse $T_p$, sample delay $T_{\rm det}$ & $170,\ 18~\mu\mathrm{s}$ \\
 Reference $T_S,T_L$ & $150,\ 250~\mu\mathrm{s}$ \\
 $s_S=s_L$, reference correlation & $1.821\times10^{-5},\ 0$ \\
 $\sqrt V,\ \tau_b,\ d$ & $4\times10^{-10},\ 500~\mathrm{s},\ 0.05$ \\
 $\Xi$ at $T_L=250,361.4~\mu\mathrm{s}$ & $4.000,\ 6.101$ \\
 \bottomrule
 \end{tabular}
\end{table}

\subsection{Optical response and readout noise}

The phase estimate shows how two dark times separate a shared phase from detuning. For the quantitative map to measurement noise, we calculate the finite-pulse response of a driven three-level $\Lambda$ system. The Lindblad equation \cite{Lindblad1976} describes equal Rabi frequencies and decay branches, with ground-state dephasing. Each combination of dark time and $\pm\pi/2$ readout phase is initialized independently, isolating the local discriminator response from inter-sequence atomic memory. The discriminator is the difference of excited-state populations at the detection delay. The corresponding experiment averages optical transmission over $1~\mu\mathrm{s}$ starting at this delay \cite{Rivera2025}.

\begin{figure}[htbp]
 \centering
 \includegraphics[width=\linewidth]{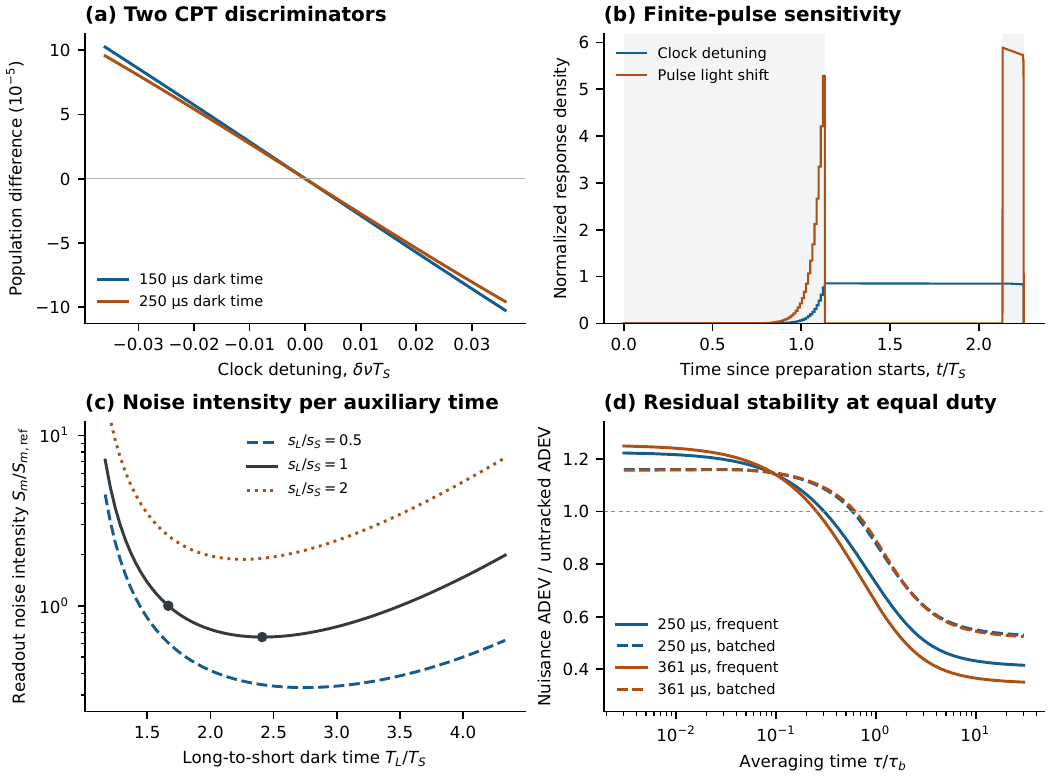}
 \caption{CPT discriminator response and equal-duty schedules for Table~\ref{tab:cpt-parameters}. (a) Population difference versus $\delta\nu T_S$ at the two reference dark times and zero pulse shift. (b) Short-channel sensitivities versus $t/T_S$, each normalized to unit integral in that coordinate. Shading marks the optical pulses. (c) Readout noise intensity $S_m/S_{m,\rm ref}$ versus $T_L/T_S$ for three long-to-short readout noise ratios $s_L/s_S$. Dots mark the two designs in (d). (d) Residual Allan deviation relative to uncorrected drift versus $\tau/\tau_b$. The solid and dashed curves use trailing boxcar observations of $1$ and $8192$ sequences, respectively, at duty $d=0.05$. Dotted curves give the matched continuous references.}
 \label{fig:ramsey}
\end{figure}

A clock detuning $\delta\nu$ acts throughout the interrogation. A pulse light shift $\nu_p$ enters the two-photon detuning during the optical pulses. Figure~\ref{fig:ramsey}(a) shows the detuning response of each discriminator at zero pulse shift. Varying the two perturbations independently gives the local Jacobian $J=\partial(D_S,D_L)/\partial(\delta\nu,\nu_p)$ by finite differences. Given the discriminator noise covariance $R_D$, inversion of $J$ yields the pulse-shift estimate in hertz. Converting to $b=g_p\nu_p$ in fractional-frequency units gives
\[
 g_p=\frac{J_{S,p}}{\nu_0 J_{S,\nu}},\qquad
 R_1=g_p^2[J^{-1}R_D(J^{-1})^T]_{22},\qquad \nu_0=\omega_0/(2\pi).
\]

Perturbations confined to short slices of the optical evolution give the two sensitivity functions. Their integrals recover the corresponding slopes of $J$. The clock response extends through the dark time, whereas the pulse-shift response is confined to the optical intervals [Fig.~\ref{fig:ramsey}(b)]. Such sensitivity profiles can also be measured experimentally \cite{Danet2014}. In the scheduling calculation, the shift varies slowly within a single sequence. A uniform window then represents the average of repeated local estimates.

We assign the eight-pattern auxiliary sequence a duration $T_{\rm seq}=8T_p+4(T_S+T_L)$. At $T_L=250~\mu\mathrm{s}$ we set the readout noise amplitude by requiring $S_{m,\rm ref}=10^{-18}~\mathrm{s}$ for fractional-frequency readout. With equal independent readout errors of fixed amplitude, the lowest $S_m$ on the sampled grid occurs at $T_L=361.4~\mu\mathrm{s}$, where $S_m=0.656S_{m,\rm ref}$. Increasing the relative noise of the long readout shifts the minimum to shorter dark times [Fig.~\ref{fig:ramsey}(c)]. Faster ground-state dephasing shortens the time over which the atomic coherence survives and also shifts the minimum to shorter dark times. The optimum is a property of the specified response and noise model.

\subsection{Comparing schedules at fixed duty}

For $M$ consecutive sequences with independent readout errors of equal variance, $R_{\rm blk}=R_1/M$. Each point on the frequent and batched curves in Fig.~\ref{fig:ramsey}(d) uses the trailing boxcar observation of Eq.~\eqref{eq:window-observation}, with $B=MT_{\rm seq}$ and $T_c=B/d$. Thus $S_m=R_{\rm blk}B=R_1T_{\rm seq}$. At fixed dark times and duty, changing $M$ alters the update interval while preserving the nominal information rate $\mathcal J=d/S_m$ and its dimensionless counterpart $\Xi=\mathcal J V\tau_b$. Changing the dark times changes both $S_m$ and $T_{\rm seq}$.

For $T_L=250~\mu\mathrm{s}$ and $361.4~\mu\mathrm{s}$, a single sequence per update gives $T_c=0.0592$ and $0.0681~\mathrm{s}$, respectively, well below the drift correlation time $\tau_b=500~\mathrm{s}$. Batches of 8192 sequences bring the intervals to $485$ and $558~\mathrm{s}$ at the same duty. At $\tau=\tau_b$, lengthening the dark time reduces the residual-to-uncorrected Allan-deviation ratio from $0.727$ to $0.652$ for frequent updates, but raises it from $0.877$ to $0.890$ for batches. The longer dark time improves the measurement information per auxiliary second in both schedules. The resulting longer update interval becomes important as it approaches the drift timescale.

\section{Discussion and conclusion}

\subsection{Interpreting a clock output}

For an output $y=b-U+n$, let $n$ denote local-oscillator and readout noise, and let $\Delta_\tau$ denote the difference of adjacent $\tau$ averages divided by $\sqrt{2}$. The Allan variance contains the residual and noise contributions and their cross covariance,
\begin{equation}
 \sigma^2_{A,y}=\sigma^2_{A,b-U}+\sigma^2_{A,n}
             +2\Cov(\Delta_\tau(b-U),\Delta_\tau n). \label{eq:clock-output}
\end{equation}
The cross covariance in Eq.~\eqref{eq:clock-output} can be nonzero when auxiliary readouts enter both the correction and the clock output. Dead time and interrogation sensitivity determine how oscillator fluctuations enter the measured frequency \cite{Santarelli1998,Danet2014}.

The results above concern the phase-averaged continuous residual. Sampling only at cycle boundaries produces a fixed-phase Allan sequence, evaluated at integer multiples of $T_c$.

\subsection{Conclusions}

At equal nominal information rate, a periodic update leaves a larger cycle-averaged error but produces smaller fluctuations between nearby averages. The value of the residual covariance at zero sets the mean-square error. Its initial slope sets the short-time Allan variance, while its area sets the long-time Allan variance. Concentrating information in periodic observations reduces the slope magnitude and increases the area for a scalar OU shift and for finite sums of independent OU modes under causal prediction.

A finite readout window changes which part of the drift enters each update and shifts the crossover in the cases studied. Holding the command changes a different quantity, the size of the correction jump, and can reverse the short-time comparison even with identical observations. For auxiliary Ramsey spectroscopy, the discriminator fixes how much information the readouts carry. Update spacing and command dynamics determine how that information affects stability at the averaging time of interest.

\printreferences

\setcounter{section}{0}
\setcounter{equation}{0}
\setcounter{figure}{0}
\setcounter{table}{0}
\renewcommand{\thesection}{S\arabic{section}}
\renewcommand{\thesubsection}{\thesection.\Alph{subsection}}
\renewcommand{\theequation}{S\arabic{equation}}
\renewcommand{\thefigure}{S\arabic{figure}}
\renewcommand{\thetable}{S\arabic{table}}
\renewcommand{\theHsection}{supp.\arabic{section}}
\renewcommand{\theHequation}{supp.\arabic{equation}}
\renewcommand{\theHfigure}{supp.\arabic{figure}}
\renewcommand{\theHtable}{supp.\arabic{table}}
\renewcommand{\citenumfont}[1]{S#1}
\renewcommand{\bibnumfmt}[1]{[S#1]}
\pdfbookmark[0]{Supplemental Material}{supplement-start}
\begin{center}
  {\Large\bfseries Supplemental Material\par}
  \vspace{4pt}
  {\large Residual Allan Stability under Periodic Auxiliary Ramsey Sampling\par}
\end{center}
\section{Exact evaluation for instantaneous updates}
\label{S:sec:point}

\begin{figure}[htbp]
 \centering
 \includegraphics[width=0.99\linewidth]{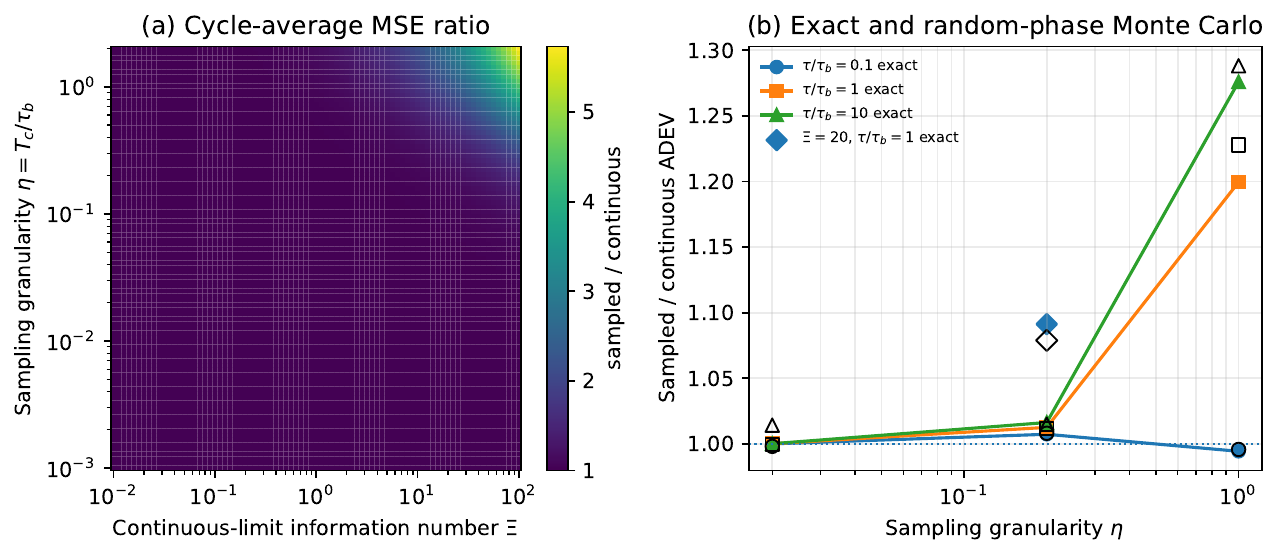}
 \caption{Instantaneous observations at matched information rate. (a) Cycle-averaged MSE relative to the continuous reference, as a function of $\Xi$ and $\eta=T_c/\tau_b$. (b) Residual Allan-deviation ratio as a function of $\eta$, at the indicated $\tau/\tau_b$. Curves show the exact covariance calculation, and symbols denote random-phase simulations.}
 \label{S:fig:sampled}
\end{figure}

In units of $\tau_b$ and $V$, let $a=e^{-\eta}$, $\alpha=1-K$, and let $p^+$ be the postupdate variance. With $r=(\Xi\eta)^{-1}$, the steady postupdate variance is the positive root of $a^2(p^+)^2+(1-a^2)(1+r)p^+-r(1-a^2)=0$. Define $\rho=a\alpha$. For a dimensionless lag $\ell=m\eta+s$, $0\leq s<\eta$, the phase-averaged covariance factors as $c_\eta(\ell)=\rho^m f(s)$ with
\begin{equation}
 f(s)=\frac{
 [\eta-(1-\alpha)s+\tfrac12(p^+-1)(1-\alpha a^2)]e^{-s}
 -\tfrac12(p^+-1)(1-\alpha)a^2e^s}{\eta}.
 \label{S:eq:covariance}
\end{equation}
Let $F_n(s)=\int_0^s t^n f(t)\,dt$ for $n=0,1$. For $x=M\eta+s$,
\begin{align}
 \mathcal C_0(x)&=F_0(\eta)\sum_{m=0}^{M-1}\rho^m+\rho^M F_0(s),\\
 \mathcal C_1(x)&=\eta F_0(\eta)\sum_{m=0}^{M-1}m\rho^m+
 F_1(\eta)\sum_{m=0}^{M-1}\rho^m+
 \rho^M[M\eta F_0(s)+F_1(s)] ,
\end{align}
where $\mathcal C_n(x)=\int_0^x\ell^n c_\eta(\ell)d\ell$. The variance of a single average is $2V[x\mathcal C_0(x)-\mathcal C_1(x)]/x^2$. Expressing the adjacent covariance via the variance of the combined $2x$ average gives
\begin{equation}
 \frac{\sigma_{A,e}^2}{V}=
 \frac{4x\mathcal C_0(x)-4\mathcal C_1(x)-2x\mathcal C_0(2x)+\mathcal C_1(2x)}{x^2}.
 \label{S:eq:sampled-allan}
\end{equation}
For the point observations, independent OU trajectories with randomly offset update phases provide a check on the Allan-deviation ratio. The simulations use 64 trajectories per case, a time step of $0.005\tau_b$ and $1400\tau_b$ per trajectory after a burn-in of $100\tau_b$. At $\tau/\tau_b=0.1,1,10$ the relative discrepancy from the analytic ratio is at most $2.4\%$ at the points shown (Fig.~\ref{S:fig:sampled}).

\section{Matched-rate ordering}
\label{S:sec:scalar-ordering}

For continuous observation at the same information rate, the scalar steady Riccati relation is $\Xi p_c^2+2p_c-2=0$ \cite{KalmanBucy1961}. Its normalized covariance is $c_c(\ell)=p_c e^{-(2/p_c-1)\ell}$. For $z=(2/p_c-1)\tau/\tau_b$, this gives the stationary OU Allan variance $Vp_c[2z-3+4e^{-z}-e^{-2z}]/z^2$ \cite{Goerlich2021}.

Between periodic updates the precision $w=1/p$ satisfies $w'=-2(w^2-w)$. The update adds $1/r=\Xi\eta$ to $w$. Cycle balance implies $\langle w^2\rangle-\langle w\rangle=\Xi/2$, where brackets denote a uniform average over cycle phase. Since $w\ge1$ is nonconstant, Jensen's inequality gives $\langle w\rangle<1/p_c$ and $\bar p=\langle 1/w\rangle>p_c$. Integrating $p'/p$ over a cycle then gives $\ln\alpha=-2\eta(\langle w\rangle-1)$.

For a lag $\ell$, the number $N$ of updates encountered has mean $\ell/\eta$. The origin-phase average of the propagated error is $c_\eta(\ell)=e^{-\ell}\langle p(u)\alpha^N\rangle$. Applying Jensen's inequality to the exponent and using $\langle\ln p\rangle\ge-\ln\langle w\rangle$ gives
\begin{equation}
 c_\eta(\ell)\ge
 \exp\!\left[-\ell+\langle\ln p\rangle+\frac{\ell}{\eta}\ln\alpha\right]
 \ge\frac{\exp[-(2\langle w\rangle-1)\ell]}{\langle w\rangle}
 >c_c(\ell).
\end{equation}
Thus the continuous covariance has the greater initial slope magnitude, $2-p_c>2-\bar p$, while the periodic covariance has the greater area. Both conclusions use the same $\Xi$.

\section{Finite-window observation}
\label{S:sec:window}

Measure time in units of $\tau_b$ and variance in units of $V$. The window width is $\beta=B/\tau_b=d\eta$ and the readout is the trailing average over $[\eta-\beta,\eta]$ plus independent noise of variance $r=(\Xi\eta)^{-1}$. Following an update of variance $p^+$ at phase zero, the covariance of future prediction errors at phases $u,v$ is
\begin{equation}
 C_0(u,v)=e^{-|u-v|}+(p^+-1)e^{-(u+v)} .
\end{equation}
Define $L=(1-e^{-\beta})/\beta$ and $H=e^{-(\eta-\beta)}L$, the conditional mean coefficient of the window average given the state at phase zero. The innovation variance $U$ and its covariance $M$ with the endpoint error are
\begin{align}
 U&=r+2(\beta-1+e^{-\beta})/\beta^2+(p^+-1)H^2,\\
 M&=L+(p^+-1)aH .
\end{align}
The steady update solves
\begin{equation}
 K=M/U,\qquad p^+=C_0(\eta,\eta)-M^2/U .
\end{equation}
The same OU increments drive the endpoint and the integrated signal, as in sampled continuous-time measurement models \cite{LjungWills2010}. Time-aggregated observations are also used for trajectory inference \cite{FoliaRattray2018}. As $\beta\to0$, $L\to1$, $H\to a$, $M\to p^-$, and $U\to p^-+r$, recovering the point update. A past phase $u$ in that cell has covariance with the updated endpoint error
\begin{equation}
 D(u)=C_0(u,\eta)-\frac{K}{\beta}\int_{\eta-\beta}^\eta C_0(u,v)\,dv .
\end{equation}
For $N\ge1$ crossed updates the propagated lag covariance is $D(u)(a-KH)^{N-1}e^{-(u+\ell-N\eta)}$, whereas for $N=0$ it is $C_0(u,u+\ell)$. Averaging these branches over $u$ and applying the Allan kernel gives the finite-window curves. The lag quadrature is split at the boundaries between these branches.

At any common $p^+\in[0,1]$, $0<M\le p^-=C_0(\eta,\eta)$. Conditional on the endpoint error, the window observation is a gain $M/p^-\le1$ times that error plus independent Gaussian noise of variance at least $r$. The instantaneous endpoint readout gives a strictly smaller posterior variance for $\beta>0$. Both scalar Riccati maps preserve order, so the window raises the steady-state posterior variance and its cycle average. The magnitude of the initial covariance slope remains $2-\bar p$ because the jump term satisfies $K M=p^--p^+$ and the OU evolution between updates is unchanged. This gives a smaller initial Allan slope than the instantaneous periodic model and hence than the continuous model. The long-time comparison for the window is evaluated from its covariance, which can include a negative one-cycle propagation factor $a-KH$.

The finite-window sweep uses $(\Xi,\eta)$ pairs $(4,0.2)$, $(4,1)$, $(4,3)$, $(0.1,1)$, $(20,1)$, $(100,1)$, and $(100,3)$. Each pair uses positive fractions $B/T_c\in\{0.01,0.05,0.1,0.25,0.5,0.75,1\}$, with the instantaneous limit as a reference. A simulation of integrated OU observations at $\Xi=4$, $\eta=1$ uses 128 stationary trajectories, 800 cycles each and 40 steps per cycle. For $B/T_c=0.25,1$ and $\tau/\tau_b=0.25,1,5$, the simulated Allan variances agree with the covariance calculation within $0.4\%$.

\section{Multiple drift timescales}
\label{S:sec:multimode}

Write $b=\bm h^T\bm b$ with $\bm h$ a vector of ones and independent OU components of stationary variances $V_i$ and correlation times $\tau_i$. We use $V=\sum_i V_i$ and $\tau_{\rm ref}=\sum_i V_i\tau_i/V$. Equal $V$ and $\tau_{\rm ref}$ match the uncorrected variance and integrated covariance across the spectra. In time units of $\tau_{\rm ref}$ and covariance units of $V$, set $A=-\operatorname{diag}(\tau_{\rm ref}/\tau_i)$, $\Sigma=\operatorname{diag}(V_i/V)$, $E(u)=e^{Au}$, $\eta=T_c/\tau_{\rm ref}$ and $\Xi=dV\tau_{\rm ref}/S_m$. Continuous sensing at the same readout budget has steady covariance $P_c$ satisfying
\begin{equation}
 0=A P_c+P_c A^T-(A\Sigma+\Sigma A^T)-\Xi P_c\bm h\bm h^T P_c .
\end{equation}
Its error propagator is $\exp[(A-\Xi P_c\bm h\bm h^T)u]$. For instantaneous periodic measurements,
\begin{align}
 P^- &=E(\eta)P^+E(\eta)^T+\Sigma-E(\eta)\Sigma E(\eta)^T,\\
 \bm K&=\frac{P^-\bm h}{\bm h^TP^-\bm h+1/(\Xi\eta)},\qquad
 P^+=(I-\bm K\bm h^T)P^-,\\
 P(u)&=\Sigma+E(u)(P^+-\Sigma)E(u)^T .
\end{align}
For $N=0$ crossed updates use $G=E(\ell)$. For $N\ge1$, $G=E(u+\ell-N\eta)[(I-\bm K\bm h^T)E(\eta)]^{N-1}(I-\bm K\bm h^T)E(\eta-u)$. The pulse-shift covariance is the cycle average of $\bm h^TP(u)G^T\bm h$. Its off-diagonal terms account for correlations in the estimation errors when the receiver observes only the sum of the modes. We use the same cycle-averaged Allan kernel as for one mode.

For the two-mode example, the fractional variances and relative times before normalizing $\tau_{\rm ref}$ are $(0.65,0.35)$ and $(0.25,4)$. For five modes they are $(0.08,0.17,0.30,0.28,0.17)$ and $(1/16,1/4,1,4,16)$. At $\Xi=4$ and $\eta=1$, the periodic-to-continuous Allan-deviation ratio at $\tau=10\tau_{\rm ref}$ is $1.276$ for one mode, $1.416$ for two modes and $1.401$ for five modes. For a cluster of nearby correlation times, with weights $(0.25,0.50,0.25)$ and relative times $(0.55,1,1.65)$ before normalization, the first detected crossing is at $0.116\tau_{\rm ref}$. In the slow-tail example the weights are $(0.96,0.04)$ and the relative times $(0.22,19.7)$. Its first crossing is at $0.0927\tau_{\rm ref}$, and its long-time Allan-variance coefficient ratio is $2.30$.

For the 64 additional spectra, the number of modes is drawn uniformly from two to seven. The weights follow a symmetric Dirichlet distribution with all concentration parameters equal to one. Relative times are drawn log-uniformly on $[1/32,32]$, rescaled to a weighted mean of one, and retained when each rescaled time lies in that interval. At $\Xi=4$, $\eta=1$, a 48-point logarithmic grid over $10^{-4}\leq \tau/\tau_{\rm ref}\leq1$ brackets sign changes. The first crossings range from $0.0442\tau_{\rm ref}$ to $0.119\tau_{\rm ref}$, with a median of $0.111\tau_{\rm ref}$. The ratios of the long-time Allan-variance coefficients range from $1.65$ to $2.59$, with a median of $1.99$.

\subsection{Matched-rate ordering for mixtures}
\label{S:sec:mixture-ordering}

For instantaneous sensing of a finite sum of independent OU modes, assume $0<\Xi,\eta<\infty$ and $0<V_i,\tau_i<\infty$. Let $c_p$ and $c_c$ be the normalized covariances of the phase-averaged periodic residual and the continuous residual at the same $\Xi$. The periodic residual gives the smaller Allan variance for sufficiently short averaging times and the larger Allan variance for sufficiently long averaging times.

\paragraph{Phase-averaged covariance.}
Keep the dimensionless units of the preceding section and put $D=-A=\operatorname{diag}(a_i)>0$, $Q=2D\Sigma$, $r=1/(\Xi\eta)$, and $\bar P=\langle P\rangle$, with brackets denoting a uniform cycle average. Write $W=P^{-1}$ and $W_c=P_c^{-1}$. Between updates, $W'=DW+WD-WQW$. The update adds $\Xi\eta\,\bm h\bm h^T$. It follows from cycle balance that, for $\mathcal F(W)=DW+WD-WQW+\Xi\bm h\bm h^T$,
\begin{equation}
 \mathcal F(\bar W)
 =\left\langle(W-\bar W)Q(W-\bar W)\right\rangle\succeq0,
 \qquad \mathcal F(W_c)=0 .
\end{equation}
Set $Z=\bar W-W_c$ and $L=D-QW_c$. For the continuous error drift $M_c=-D-\Xi P_c\bm h\bm h^T$, the Riccati equation gives $M_cP_c+P_cM_c^T=-Q-\Xi P_c\bm h\bm h^TP_c\prec0$. Hence $M_c$ and the similar matrix $L=P_c M_c^T P_c^{-1}$ are Hurwitz. Expanding around $W_c$ gives
\begin{equation}
 L^TZ+ZL=\mathcal F(\bar W)+ZQZ\succeq0
 \quad\Longrightarrow\quad
 Z\preceq0 .
\end{equation}
The implication follows from $Z=-\int_0^\infty e^{L^Tt}[\mathcal F(\bar W)+ZQZ]e^{Lt}dt$. Jensen's inequality for matrix inversion gives $\bar P\succeq\bar W^{-1}\succeq P_c$. The first inequality is strict in the measured direction. Writing $B_0=\bar W^{-1}$, the identity $\bar P-B_0=\langle(P-B_0)W(P-B_0)\rangle$ follows from $PW=WP=I$. For $\bm v=\bm h$, $P(u)\bm h$ changes over the cycle because the update reduces the variance of the measured sum.

\paragraph{Short-time coefficient.}
Let $\bm q=P^-\bm h$, $s=r+\bm h^T\bm q$ be the innovation variance, $\tilde{\bm q}=\bm q/\sqrt{s/r}$, and $\bm q_c=P_c\bm h$. The discrete covariance drop is $P^--P^+=\bm q\bm q^T/s$. Integrating $P'=-DP-PD+Q$ over the cycle and using the continuous Riccati equation gives
\begin{equation}
 \bar P_{ii}=\Sigma_{ii}-\frac{\Xi\,\tilde q_i^2}{2a_i},
 \qquad
 [P_c]_{ii}=\Sigma_{ii}-\frac{\Xi\,q_{c,i}^2}{2a_i}.
 \label{S:eq:mixture-componentwise}
\end{equation}
The covariance order yields $|\tilde q_i|\leq |q_{c,i}|$, with strict inequality for at least one component. All $q_{c,i}$ are positive. To see this, let $\bm K_c=\Xi P_c\bm h$. The error is orthogonal to the current estimate, so \(q_{c,i}\) is the cross covariance of state component $i$ with the error of the measured sum. The causal filter and $\operatorname{Cov}[b_i(t),b(t-u)]=\Sigma_{ii}e^{-a_i u}$ give
\begin{equation}
 \frac{q_{c,i}}{\Sigma_{ii}}
 =1-\bm h^T(a_i I+D+\bm K_c\bm h^T)^{-1}\bm K_c
 =\frac{\det(a_i I+D)}{\det(a_i I+D+\bm K_c\bm h^T)}>0 .
\end{equation}
Both determinants are positive because $a_i>0$ and the error drift is stable.

For positive infinitesimal lag, the periodic covariance derivative has a drift term $-\bm h^TD\bar P\bm h$ and an update term $-\bm h^T(P^--P^+)\bm h/\eta$. Eliminating the drift term using the cycle balance, and doing the same with the continuous Riccati equation, yields
\begin{equation}
 -c_p'(0^+)=\tfrac12\bm h^TQ\bm h+\tfrac{\Xi}{2}(\bm h^T\tilde{\bm q})^2,
 \qquad
 -c_c'(0^+)=\tfrac12\bm h^TQ\bm h+\tfrac{\Xi}{2}(\bm h^T\bm q_c)^2 .
\end{equation}
Since $0<\bm h^T\tilde{\bm q}<\bm h^T\bm q_c$, the periodic cusp has smaller magnitude. The leading normalized Allan variance for $x=\tau/\tau_{\rm ref}\to0$ is $-\tfrac23 c'(0^+)x$.

\paragraph{Long-time coefficient.}
The long-time coefficient is the two-sided covariance integral, or the spectrum at zero in the convention $S(0)=\int_{\mathbb R} c(u)\,du$. Define the uncorrected continuous spectrum at zero and the corresponding scaled point-sample spectrum by
\begin{equation}
 S_0=2\sum_i \frac{\Sigma_{ii}}{a_i},\qquad
 S_a=\eta\sum_i \Sigma_{ii}\frac{1+f_i}{1-f_i}>S_0,
 \qquad f_i=e^{-a_i\eta}.
\end{equation}
The inequality uses $\coth z>1/z$ for $z>0$. Let $g_p$ be the zero-frequency gain from the readout sequence to the cycle averages of the periodic estimate, and let $g_c$ be the corresponding gain for the continuous filter. Period averaging of the raw process gives both its discrete auto-spectrum at zero and its cross-spectrum with the raw point samples equal to $S_0/\eta$, by summing the covariances across all lags. The noisy readout has spectrum $(S_a+1/\Xi)/\eta$ at zero. For cycle averages $\bar e_j=\eta^{-1}\int_{j\eta}^{(j+1)\eta}e(t)dt$ in normalized time, tiling the real line with the second integration window gives $\eta\sum_{j\in\mathbb Z}\Cov(\bar e_0,\bar e_j)=\int_{\mathbb R} c_p(u)du$, in normalized covariance units. The long-time coefficients are
\begin{equation}
 L_p=S_0(1-2g_p)+(S_a+1/\Xi)g_p^2,\qquad
 L_c=S_0(1-2g_c)+(S_0+1/\Xi)g_c^2 .
 \label{S:eq:mixture-dc}
\end{equation}

For the periodic estimate set $F=E(\eta)$, $B=(I-\bm K\bm h^T)F$, $\bm K=\bm q/s$, and $v=\bm h^TF(I-F)^{-1}\bm K$. The stationary posterior covariance obeys
\[
 P^+-BP^+B^T=(I-\bm K\bm h^T)(\Sigma-F\Sigma F^T)(I-\bm K\bm h^T)^T+r\bm K\bm K^T\succ0 .
\]
Here $I-\bm K\bm h^T$ is invertible since $r>0$, so $B$ is Schur stable. The innovations $\nu_k=z_k-\bm h^TF\widehat{\bm b}_{k-1}^+$ generate $\widehat{\bm b}_k^+=F\widehat{\bm b}_{k-1}^++\bm K\nu_k$. The transfer from innovations to $z_k$ has gain $1+v$ at zero frequency, hence $S_a+1/\Xi=\eta s(1+v)^2$. Averaging the predicted state over a cycle gives
\begin{align}
 g_p
 &=\frac{\bm h^TD^{-1}(I-F)(I-B)^{-1}\bm K}{\eta}
 =\frac{\bm h^TD^{-1}\bm q}{\eta s(1+v)}
 =\frac{\Xi\,\bm h^TD^{-1}\tilde{\bm q}}{\sqrt{1+\Xi S_a}} .
\end{align}
Here $1+v>0$, as it equals $\det(I-B)/\det(I-F)$ with both $B$ and $F$ stable. Also $q_i>0$. Indeed, the innovation whitening filter is $\det(I-F z)/\det(I-B z)$ in the lag variable $z$. The prior estimation error is orthogonal to the predictor, so $q_i=\Cov(b_i(t_k),\nu_k)$ in normalized units. This covariance evaluates the whitening transfer at $z=f_i$, since the state-to-past-readout covariance at lag $j$ is $\Sigma_{ii}f_i^j$. The determinant ratio is positive on $0<z<1$ by stability.

For the continuous filter, let $C_c=\bm h^TD^{-1}\bm q_c$. Multiplying its Riccati equation by $\bm h^TD^{-1}$ and $D^{-1}\bm h$ gives $S_0=2C_c+\Xi C_c^2$. The continuous zero-frequency gain is therefore
\begin{equation}
 g_c=\frac{\Xi C_c}{1+\Xi C_c}
 =\frac{\Xi C_c}{\sqrt{1+\Xi S_0}}
 =1-\frac{1}{\sqrt{1+\Xi S_0}} .
\end{equation}
Equation~\eqref{S:eq:mixture-componentwise} and $S_a>S_0$ imply $0<g_p<g_c<S_0/(S_0+1/\Xi)$. The quadratic $S_0(1-2g)+(S_0+1/\Xi)g^2$ strictly decreases over this gain range. Equation~\eqref{S:eq:mixture-dc} then gives $L_p>L_c$. The opposite short- and long-time orderings imply at least one crossing by continuity.

\section{Calibrating the auxiliary block variance}
\label{S:sec:cpt}

A three-level master equation \cite{Lindblad1976} provides model discriminator slopes for coherent population trapping. During an optical pulse of phase $\theta$ the Hamiltonian is
\begin{equation}
 H/\hbar=
 \begin{pmatrix}
 -(\delta+\Delta_{\rm LS})/2 & 0 & \Omega/2\\
 0 & (\delta+\Delta_{\rm LS})/2 & \Omega e^{-i\theta}/2\\
 \Omega/2 & \Omega e^{i\theta}/2 & -\Delta
 \end{pmatrix}.
\end{equation}
During the dark time set $\Omega=\Delta_{\rm LS}=0$. Equal excited-state branching rates are $\Gamma_e/2$, and the ground dephasing jump is $\sqrt{\gamma_g/2}\operatorname{diag}(1,-1,0)$. Use $\Gamma_e/(2\pi)=4.6$ MHz, $\Omega/(2\pi)=0.30$ MHz, $\gamma_g/(2\pi)=0.80$ kHz, $\Delta=0$, and a $170~\mu$s preparation pulse from an incoherent equal mixture of the ground states. The error signals are differences of excited-state populations $18~\mu$s into the two $\theta=\pm\pi/2$ readout branches. The reference dark times are $T_S=150~\mu$s and $T_L=250~\mu$s, values used in a symmetric Ramsey-CPT implementation \cite{Rivera2025}.

Finite differences with respect to $\delta/(2\pi)$ and $\Delta_{\rm LS}/(2\pi)$ give the short/long readout response $J$. At the reference dark times, the clock and pulse-shift columns of $J$ are $(-4.313,-4.095)^T\times10^{-7}~\mathrm{Hz}^{-1}$ and $(-6.281,-3.800)^T\times10^{-8}~\mathrm{Hz}^{-1}$, respectively. For readout noise covariance $R$, the second diagonal element of $(J^TR^{-1}J)^{-1}$ is the noise variance of the inferred light shift in squared hertz. Conversion to the effective fractional-frequency shift uses the calibrated light-shift-to-clock coefficient. Multiplying the resulting fractional variance by the block duration gives the white-noise intensity $S_m$ for the tracking model. For equal, independent short- and long-channel errors, the noise covariance is proportional to $I$. Its common scale fixes the absolute intensity and leaves the minimizing dark time unchanged. Over $175\leq T_L/(\mu\mathrm{s})\leq650$, the minimum occurs at $361.4~\mu\mathrm{s}$. Increasing $s_L/s_S$ from $0.5$ to $2$ shifts the minimum from $409.5$ to $337.3~\mu\mathrm{s}$. At equal readout noise, increasing the ground-state dephasing rate $\gamma_g/(2\pi)$ from $0.4$ to $1.2~\mathrm{kHz}$ moves it from $541.8$ to $295.3~\mu\mathrm{s}$.

\section{Held-command covariance and simulation}
\label{S:sec:hold}

Use normalized time and frequency, and let $\bm x_k=(b_k,\widehat b_k^+)^T$ at the update. Orthogonality of the conditional estimate and its error gives the stationary covariance $\Sigma_x$ and transition $F$,
\begin{equation}
 \Sigma_x=\begin{pmatrix}1&1-p^+\\1-p^+&1-p^+\end{pmatrix},\qquad
 F=\begin{pmatrix}a&0\\aK&a(1-K)\end{pmatrix}.
\end{equation}
The driving covariance is $(1-a^2)\bm g\bm g^T+rK^2\bm e_2\bm e_2^T$, with $\bm g=(1,K)^T$ and $\bm e_2=(0,1)^T$. The OU increment from the update to phase $u$ is $\nu_u$, independent of $\bm x_k$, with variance $1-e^{-2u}$. Let $q(u)=(e^{-u},-1)$ be a row vector. The held-command residual at phase $u$ is $q(u)\bm x_k+\nu_u$.

For a lag $\ell$, let $n=\lfloor(u+\ell)/\eta\rfloor$ and $v=u+\ell-n\eta$. Conditional on phase $u$, the normalized covariance is
\begin{equation}
 C_H(u,\ell)=
 \begin{cases}
 q(u)\Sigma_x q(v)^T+(1-e^{-2u})e^{-\ell}, & n=0,\\
 \big[q(u)\Sigma_x F^T+(1-e^{-2u})e^{u-\eta}\bm g^T\big]
      (F^T)^{n-1}q(v)^T, & n\ge1 .
 \end{cases}
\end{equation}
The within-cycle process noise appears in both residuals when $n=0$. For $n\ge1$, its cross covariance with the next endpoint innovation is $(1-e^{-2u})e^{u-\eta}\bm g^T$. Subsequent process and measurement increments are independent of the first residual. Integrating over uniform $u$ and then applying the Allan kernel gives the held-command curves. The phase integral is split at each change in $n$.

The simulation propagates the state and its integrated value using the joint exact OU transition over a step $h$. Conditional on the initial state, the endpoint and integral innovation have variances $1-e^{-2h}$ and $2h-4(1-e^{-h})+1-e^{-2h}$, and covariance $(1-e^{-h})^2$. The predicted and held command integrals are exact on those steps. We simulate the continuous reference from its steady OU residual at variance $p_c$ and rate $2/p_c-1$ using the same exact joint transition, rescaled to those parameters. Adjacent block averages start on a uniform grid of phases, using 20 and 40 phase steps per cycle. Each run uses 160 independent trajectories and 1200 retained cycles per trajectory after 60 burn-in cycles. At $\Xi=4$, $\eta=1$, we compare the simulated and analytic Allan variances for the predicted command, the held command and the continuous reference at $\tau/\tau_b=0.25,1,5$. Over these nine comparisons, the largest relative discrepancies are $0.83\%$ at 20 steps and $0.96\%$ at 40 steps. They include finite-ensemble variation. In Fig.~\ref{fig:windows}(c), each simulation point and its interval are divided by the analytic continuous-reference Allan variance. The intervals extend $1.96$ trajectory standard errors on either side of the 40-step estimates.

\renewcommand{\refname}{Supplemental references}
\printreferences

\end{document}